\documentclass[12pt]{article}
\usepackage{amssymb}
\usepackage{amsmath}
\usepackage{amsthm}
\usepackage{bm} % used for get math symbols in bold use \bm
\usepackage{graphics}
\usepackage{graphicx}
\usepackage[T1]{fontenc}
\usepackage{longtable}
\usepackage{natbib}
\usepackage{setspace}
\usepackage{times}

\usepackage{epsfig}
\usepackage{fancyvrb}
\usepackage{multirow}
\usepackage[affil-it]{authblk}
\usepackage{verbatim}
\usepackage{tablefootnote}
\usepackage[explicit]{titlesec}
\usepackage{dcolumn}

\usepackage{subcaption}

\usepackage{tikz}
\usetikzlibrary{shapes,snakes, graphs}

\usetikzlibrary{arrows,%
                petri,%
                topaths}%
\usepackage{tkz-berge}
\usepackage{pgf}
\usepackage{tikz}
\usetikzlibrary{arrows,automata}
\usetikzlibrary{decorations.markings}

\newtheorem{theorem}{Theorem}
\newtheorem*{theorem*}{Theorem}

\def\ci{\mbox{\ensuremath{\perp\!\!\!\perp}}}
\def\E{\mathbb{E}}

\titleformat{\section}
  {\normalfont\Large\bfseries\centering}{\thesection.}{1em}{\MakeUppercase{#1}}
  
\begin{document}

\allowdisplaybreaks

\def\spacingset#1{\renewcommand{\baselinestretch}%
{#1}\small\normalsize} \spacingset{1}

\title{\bf On semiparametric estimation of a path-specific effect in the presence of mediator-outcome confounding}
\author{Caleb H. Miles, Ilya Shpitser, Phyllis Kanki, Seema Meloni, and Eric J. Tchetgen Tchetgen\thanks{Caleb H. Miles is Postdoctoral Fellow, Division of Biostatistics, University of California, Berkeley, Berkeley, CA 94720-7358. Ilya Shpitser is John C. Malone Assistant Professor, Department of Computer Science, Johns Hopkins University, Baltimore, MD 21218-2608. Phyllis Kanki is Professor and Seema Meloni is Research Scientist, Department of Immunology and Infectious Diseases, Harvard T.H. Chan School of Public Health, Boston, MA 02115. Eric J. Tchetgen Tchetgen is Professor, Departments of Biostatistics and Epidemiology, Harvard T.H. Chan School of Public Health, Boston, MA 02115. The authors gratefully acknowledge the hard work and dedication of the clinical, data, and laboratory staff at the PEPFAR supported Harvard/AIDS Prevention Initiative in Nigeria (APIN) hospitals that provided secondary data for this analysis. This work was funded, in part, by the US Department of Health and Human Services, Health Resources and Services Administration (U51HA02522) and by the National Institutes of Health (R01AI104459-01A1). The contents are solely the responsibility of the authors and do not represent the official views of the funding institutions.}\hspace{.2cm}}
\date{}
\maketitle
\bigskip

\begin{abstract}
\noindent Path-specific effects are a broad class of mediated effects from an exposure to an outcome via one or more causal pathways with respect to some subset of intermediate variables. The majority of the literature concerning estimation of mediated effects has focused on parametric models with stringent assumptions regarding unmeasured confounding. We consider semiparametric inference of a path-specific effect when these assumptions are relaxed. In particular, we develop a suite of semiparametric estimators for the effect along a pathway through a mediator, but not some exposure-induced confounder of that mediator. These estimators have different robustness properties, as each depends on different parts of the observed data likelihood. One of our estimators may be viewed as combining the others, because it is locally semiparametric efficient and multiply robust. The latter property is illustrated in a simulation study. We apply our methodology to an HIV study, in which we estimate the effect comparing two drug treatments on a patient's average log CD4 count mediated by the patient's level of adherence, but not by previous experience of toxicity, which is clearly affected by which treatment the patient is assigned to, and may confound the effect of the patient's level of adherence on their virologic outcome.
\end{abstract}

\noindent%
\vfill

\newpage
\spacingset{1.45} % DON'T change the spacing!

\section{Introduction}
A literature within causal inference has recently emerged concerning the definition, identification, and estimation of direct and indirect effects in fully nonparametric settings, which includes settings where certain interactions and non-linearities may be present  \citep{robins1992identifiability,robins1999testing,robins2003semantics,pearl2001direct,
avin2005identifiability,vanderweele2009conceptual,vanderweele2010odds,imai2010general,
imai2010identification,tchetgen2012semiparametric,tchetgen2014estimation}. This strand of work is based on ideas developed by \cite{robins1992identifiability} and \cite{pearl2001direct}, and uses the language of potential outcomes \citep{rubin1974estimating,splawa1990application} to give a nonparametric definition of effects involved in mediation analysis.

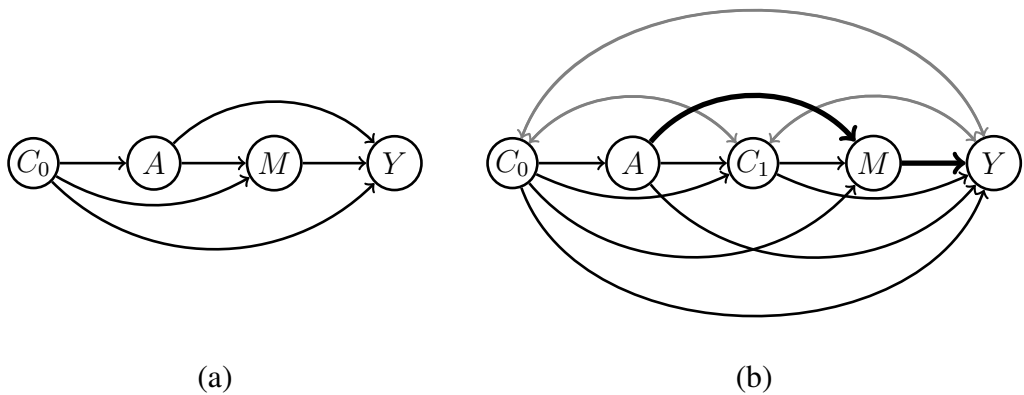
\begin{figure}[h]
\centering
\begin{tabular}{ccc}
\begin{tikzpicture}[baseline={(A)}, scale=.8, ->, line width=1pt]
\tikzstyle{every state}=[draw=none]
\node[shape=circle, draw, inner sep=.3mm] (A) at (0,0) {$C_0$};
\node[shape=circle, draw, inner sep=1mm] (B) at (2,0) {$A$};
\node[shape=circle, draw, inner sep=.7mm] (D) at (4,0) {$M$};
\node[shape=circle, draw, inner sep=1mm] (E) at (6,0) {$Y$};

  \path 	(A) edge (B)
			(A) edge  [bend right=30] (D)
			(A) edge  [bend right=45] (E)
			(D) edge (E)
			(B) edge (D)
			(B) edge  [bend left=45] (E)
					  ;
\end{tikzpicture}
& 
&
\begin{tikzpicture}[baseline={(A)}, scale=.8, ->, line width=1pt]
\tikzstyle{every state}=[draw=none]
\node[shape=circle, draw, inner sep=.3mm] (A) at (-2,0) {$C_0$};
\node[shape=circle, draw, inner sep=1mm] (B) at (0,0) {$A$};
\node[shape=circle, draw, inner sep=.3mm] (C) at (2,0) {$C_1$};
\node[shape=circle, draw, inner sep=.7mm] (D) at (4,0) {$M$};
\node[shape=circle, draw, inner sep=1mm] (E) at (6,0) {$Y$};

  \path 	(E) edge [color=gray, bend right=50]  (C)
			(C) edge [color=gray, bend left=50]  (E)
			(A) edge [color=gray, bend left=50] (C)
			(C) edge [color=gray, bend right=50] (A)
			(A) edge [color=gray, bend left=70] (E)
			(E) edge [color=gray, bend right=70] (A)
  			(A) edge (B)
			(A) edge  [bend right=25] (C)
			(A) edge  [bend right=50] (D)
			(A) edge  [bend right=70] (E)
			(B) edge (C)
			(C) edge (D)
			(D) edge [line width=2pt] (E)
			(B) edge  [bend left=50, line width=2pt] (D)
			(B) edge  [bend right=50] (E)
			(C) edge  [bend right=25] (E)
					  ;
\end{tikzpicture}
\\
(a)	&	&	(b)
\end{tabular}
\caption{(a) The standard mediation graph with a single intermediate variable, $M$. (b) A causal graph with unobserved confounding, and two intermediate variables, one of which ($C_1$) confounds the effect of the other ($M$) on $Y$.\label{fig:3node}}
\end{figure}

Path-specific effects belong to a large class of mediated effects that capture the effect of an exposure, $A$, on a post-treatment outcome, $Y$, through one or more causal pathways, which involve some subset of intermediate variables. The simplest and most traditional mediation setting arises when causal pathways are considered with respect to a single intermediate variable, say $M$, as depicted in the directed acyclic graph in Fig.~\ref{fig:3node}.a. 
The causal pathway through $M$ in this graph is known as the natural indirect effect, and the causal pathway not through $M$ is known as the pure (or natural) direct effect. Identification and inference corresponding to causal mediation queries in this graph is for the most part resolved \citep{robins1992identifiability,pearl2001direct,petersen2006estimation,van2008direct,
robins2010alternative,tchetgen2012semiparametric,
tchetgen2014estimation}.

More recently, causal mediation analysis grounded in the counterfactual framework has considered tools for other mediated effects of interest. In the present work, we are concerned with the considerably more challenging setting in Fig.~\ref{fig:3node}.b, in which there is another intermediate variable, $C_1$, subsequent to $A$, but occurring before $M$ and $Y$. This variable may be affected directly by $A$, and may in turn confound the effect of $M$ on $Y$, which has been shown to render certain mediated effects nonparametrically unidentifiable \citep{shpitser2013counterfactual}. Further, the presence of the gray bi-directed edges between $C_0$, $C_1$, and $Y$ represent potential unmeasured confounding. We are interested in estimating the path-specific effect along the emboldened pathway $A\rightarrow M\rightarrow Y$, which we term $\mathcal{P}_{AMY}$. This effect captures the effect of $A$ on $Y$ mediated by $M$ through mechanisms by which $A$ affects $M$ other than those governed by $C_1$. In \cite{doi:10.1080/01621459.2017.1295862}, we considered an HIV example in which $\mathcal{P}_{AMY}$ is of substantive interest. There, $A$ was an indicator of assignment to one first-line antiretroviral therapy treatment vs.~another, $C_1$ was a measure of adherence and toxicity over the next six months, $M$ was adherence over the subsequent six months, and $Y$ was an indicator of virological failure. Here, we will consider the same setting with CD4 count as outcome.

\cite{avin2005identifiability} and \cite{shpitser2013counterfactual} have considered general identification conditions for path-specific effects. In \cite{doi:10.1080/01621459.2017.1295862}, we presented conditions for identification in the presence of unmeasured confounding, and developed a maximum likelihood estimator of this effect. In this paper, we build on this previous work, and develop semiparametric estimation theory for this effect, allowing for other parts of the likelihood to remain unrestricted. We derive the efficient influence function of the identifying functional for $\mathcal{P}_{AMY}$, and present a suite of semiparametric estimators with different robustness properties, as each depends on different parts of the observed data likelihood. One of these may be viewed as combining the others, because it is locally semiparametric efficient and multiply robust. By multiply robust, we mean that while this estimator depends on estimation of four nuisance parameters, it remains consistent and asymptotically normal provided only one of three possible subsets of these nuisance parameters are consistently estimated.

\section{Definition and identification formula}

Suppose that one has observed independent and identically distributed realizations of $O = (C_0,\allowbreak A,\allowbreak C_1,\allowbreak M,\allowbreak Y)$. In the HIV application, $A$ indicates the exposure to one of two first-line anti-retroviral therapy treatments prescribed to most HIV patients in Nigeria. For notational convenience, we will use $A=a'$ to denote the reference level treatment and $A=a$ to denote the comparison level treatment. The variable $C_1$ contains a measure of toxicity due to exposure to a particular treatment regimen, ascertained six months after treatment initiation, $M$ is the adherence level during the subsequent six months, $Y$ is the patient's log CD4 count at the end of the year, and $C_0$ is the vector of all baseline covariates. Letting $\mathrm{supp}(\cdot)$ denote the support of its argument, we make the following positivity assumptions \citep{robins1986new}: $\inf_{m\in\mathrm{supp}(M)} f_{M\mid C_1,A,C_0}(m\mid C_1,A,C_0)>0$ almost everywhere, $\inf_{c_1\in\mathrm{supp}(C_1)} f_{C_1\mid A,C_0}(c_1\mid A,C_0)>0$ almost everywhere, and $0<f_{A\mid C_0}(a\mid C_0)<1$ almost everywhere.

To formalize the path-specific effect of interest, we now introduce counterfactuals (or potential outcomes). Let $Y(a^*)$ denote a subject's outcome if treatment $A$ were set, possibly contrary to fact, to $a^*\in\{a',a\}$. In the context of mediation, there will also be potential outcomes for the intermediate variables. We define $C_1(a^*)$, $M(a^*)$, $M(a^*,c_1)$, $Y(a^*)$, $Y(a^*,m)$, and $Y(a^*,c_1,m)$ analogously. We adopt a standard set of consistency assumptions, which link these counterfactuals to the observed variables. Generally, for a set of variables $W_1$ and a variable $W_2$ for which the counterfactual $W_2(w_1)$ is defined, if $W_1=w_1$, then $W_2(w_1)=W_2$ almost everywhere.

We further assume that nested counterfactuals are well defined. 
\begin{comment}
One example of a nested counterfactual is $Y\{a,M(a')\}$, which can be used to decompose the average causal effect, yielding the pure direct effect and natural indirect effect.
\begin{align*}
&E\left\{ Y(a)\right\} -E\left\{ Y(a')\right\} \\
&=\overset{\mathrm{total\text{ }effect}}{\overbrace{E\left[
Y\{a,M(a)\}\right] -E\left[ Y\{a',M(a)\}\right] }} \\
&=\overset{\mathrm{natural\text{ }indirect\text{ }effect}}{\overbrace{E%
\left[ Y\{a,M(a)\}\right] -E\left[ Y\{a,M(a')\}\right] }}+\overset{%
\mathrm{pure\text{ }direct\text{ }effect}}{\overbrace{E\left[
Y\{a,M(a')\}\right] -E\left[ Y\{a',M(a')\}\right] }}.\newline
\end{align*}
\end{comment}
The path-specific effect of $A$ on $Y$ along the path $A\rightarrow M\rightarrow Y$, with respect to comparison treatment value $a$ and baseline value $a'$ on the mean difference scale is formally defined in terms of the difference in expectations of two nested counterfactuals:
\[\mathcal{P}_{AMY}\equiv E(Y[M\{a,C_1(a')\},C_1(a'),a']) - E(Y[M\{a',C_1(a')\},C_1(a'),a']).\]
The second term reduces to $E\{Y(a')\}$, i.e., the average outcome a patient would have experienced had they been assigned the baseline-value exposure. This term is identified under the no unobserved confounding condition, $Y(a')\ci A\mid C_0$, which holds under both graphs in Fig.~\ref{fig:3node}, and its estimation has been studied extensively (see \cite{rubin1978bayesian,rosenbaum1983central,rosenbaum1984conditional,
robins1992estimating}). As such, the remainder of our discussion will focus on the first term, which we denote $\beta_0\equiv E(Y[M\{a,C_1(a')\},C_1(a'),a'])$. This term defines the average outcome a patient would experience under an intervention which assigns the patient to the reference-level exposure $a'$, while experiencing the toxicity associated with this exposure, $C_1(a')$, however with adherence profile associated with the comparison-level exposure and toxicity profile associated with reference-level treatment, $M\{a,C_1(a')\}$.

In general, it is possible to give an inductive definition of a path-specific effect of $A$ on $Y$ for an arbitrary bundle of pathways, which results in a quantity that is a function of a nested potential outcome. A general definition for the static treatment and single outcome case is given in \cite{pearl2001direct} and \cite{avin2005identifiability}. \cite{shpitser2013counterfactual} extends this definition to longitudinal settings with repeated exposures and mediators.

Theorem 1 in \cite{doi:10.1080/01621459.2017.1295862} (a special case of Theorem 4 in \cite{shpitser2013counterfactual}) gives the nonparametric identification formula
\begin{align}
\beta_0=\iiint\limits_{c_0,c_1,m} E(Y\mid m,c_1,a',c_0)dF(m\mid c_1,a,c_0)dF(c_1\mid a',c_0)dF(c_0)
\end{align}
under independences which hold in a relaxation of the Markovian model represented by the graph in Fig.~\ref{fig:3node}.b, i.e., in a model where there is no unmeasured confounding of the exposure-outcome and mediator-outcome relationships. A detailed discussion of the causal assumptions implied by this model can be found in that paper.

\section{Maximum likelihood estimation}
Thus far, we have considered identification under a nonparametric statistical model, $\mathcal{M}_{np}$, for the observed data, making our identifying functional of $\mathcal{P}_{AMY}$ valid under any possible model for the data satisfying the given positivity conditions. However, because inference in $\mathcal{M}_{np}$ is often impractical in situations with numerous confounders $(C_0, C_1)$ relative to sample size \citep{robins1997toward}, we will often be unable to estimate $\mathcal{P}_{AMY}$ nonparametrically, and instead must posit parametric models.

We now consider the first of our four estimators for $\beta_0$. By considering the identifying functional (1) as four nested expectations, it is clear that we can fit three appropriate regression models with parameters $\gamma_1$, $\gamma_2$, and $\gamma_3$ using maximum likelihood, and obtain a substitution estimator by plugging the predicted means under these models into the functional; the outermost mean can then be estimated empirically. Thus, the maximum likelihood estimator is
\[\hat{\beta}_{mle} \equiv \mathbb{P}_n\left(\hat{ E}\left[\hat{ E}\left\{\hat{ E}\left(Y\mid M,C_1,a',C_0;\hat{\gamma}_
1\right)\mid C_1,a,C_0;\hat{\gamma}_2\right\}\mid a',C_0;\hat{\gamma}_3\right]\right),\]
where $\mathbb{P}_n$ denotes the empirical mean. Inference can be conducted using standard maximum likelihood theory. This estimator is guaranteed to be consistent only under correct specification of these three models, and can be useful if the propensity score is deemed difficult to model. See \cite{doi:10.1080/01621459.2017.1295862} for additional details on maximum likelihood-based inference for $\beta$.

\section{Semiparametric inference}
\subsection{Two semiparametric estimators}
Define $ M^{\rm ratio}(\allowbreak M,\allowbreak C_1,\allowbreak C_0) \allowbreak  \equiv \allowbreak  f( \allowbreak M\mid  \allowbreak C_1, \allowbreak a, \allowbreak C_0) \allowbreak / \allowbreak f( \allowbreak M\mid  \allowbreak C_1, \allowbreak a', \allowbreak C_0)$, $C_1^{\rm ratio}(\allowbreak C_1,\allowbreak C_0) \allowbreak  \equiv \allowbreak  f( \allowbreak C_1\mid  \allowbreak a, \allowbreak C_0) \allowbreak / \allowbreak f( \allowbreak C_1\mid  \allowbreak a', \allowbreak C_0)$, and $1_{x}(\cdot)$ to be the indicator function. We consider two estimators based on alternative representations of (1), as shown in the supplementary materials:
\begin{align*}
\hat{\beta}_a &\equiv \mathbb{P}_n\left\{\frac{1_{a'}(A)}{\hat{f}(a'\mid C_0)}\hat{M}^{\rm ratio}Y\right\}\\
\hat{\beta}_b &\equiv \mathbb{P}_n\left\{\frac{1_a(A)}{\hat{f}(a\mid C_0)}(\hat{C}_1^{\rm ratio})^{-1}\hat{ E}(Y\mid M,C_1,a',C_0)\right\}.
\end{align*}

Similar to the maximum likelihood estimator, these estimators also involve plugging in estimated regression models. When $C_1$ or $M$ are continuous, one can avoid estimating their conditional densities by instead estimating their conditional density ratios directly. The conditional density ratios can in fact be estimated using regression models for the exposure because by Bayes' theorem,
\[C_1^{\rm ratio}(\allowbreak C_1,\allowbreak C_0) \allowbreak  \equiv\frac{f(C_1\mid a,C_0)}{f(C_1\mid a',C_0)} = \frac{f(a\mid C_1,C_0)}{f(a'\mid C_1,C_0)}\times \frac{f(a'\mid C_0)}{f(a\mid C_0)}\]
and
\[M^{\rm ratio}(\allowbreak M,\allowbreak C_1,\allowbreak C_0) \allowbreak  \equiv\frac{f(M\mid a,C_1,C_0)}{f(M\mid a',C_1,C_0)} = \frac{f(a\mid M,C_1,C_0)}{f(a'\mid M,C_1,C_0)}\times\frac{f(a'\mid C_1,C_0)}{f(a\mid C_1,C_0)}.\]
The parameters $f(a\mid C_0)$, $f(a\mid C_1, C_0)$, and $f(a\mid M,C_1,C_0)$ are not variationally independent. For instance, when $A\ci C_1\mid C_0$, $f(a\mid C_1, C_0)$ is restricted to be equivalent to $f(a\mid C_0)$. One can ensure compatibility between these models by the following procedure. First, specify a logistic model for $f(a\mid C_0)$, then specify a model for $f(C_1\mid A,C_0)$ such that $\log\{f(C_1\mid a,C_0)/f(C_1\mid a',C_0)\}$ is linear in functions of $C_0$ and $C_1$ not depending on unknown parameters of the model for $f(C_1\mid A,C_0)$. For example, for $C_1$ one could use the model $C_1\mid A,C_0\sim N(\alpha_0+\alpha_1A+\alpha_2C_0,\sigma^2)$, since \[\log\frac{f(C_1\mid a,C_0)}{f(C_1\mid a',C_0)}=\alpha_2(C_1-\alpha_0-\alpha_1C_0)(a-a')-\frac{\alpha^2_2}{2}(a^2-a'^2),\]
which is linear in $C_0$ and $C_1$. Then a compatible logistic model for $f(a\mid C_1,C_0)$ can be obtained by taking as regressors the union of the linear terms in the logistic model for $f(a\mid C_0)$ and $\log\{f(M\mid a,C_1,C_0)/f(M\mid a',C_1,C_0)\}$. For example, if the logistic model for $f(a\mid C_0)$ has the linear component $\delta_0+\delta_1C_0$, and the above normal model is used for $C_1$, then
\begin{align*}
\log\frac{f(a\mid C_1,C_0)}{f(a'\mid C_1,C_0)}&=\log\frac{f(C_1\mid a,C_0)}{f(C_1\mid a',C_0)}+\log\frac{f(a\mid C_0)}{f(a'\mid C_0)}\\
&= \alpha_2(C_1-\alpha_0-\alpha_1C_0)(a-a')-\frac{\alpha^2_2}{2}(a^2-a'^2)+\delta_0+\delta_1C_0\\
&= \zeta_0+\zeta_1C_0+\zeta_2C_1
\end{align*}
for an appropriate choice of $\zeta$, and hence this logistic model for $f(a\mid C_1,C_0)$ will be compatible with $f(a\mid C_0)$. A compatible model for $f(a\mid M,C_1,C_0)$ can be obtained analogously using the resulting model for $f(a\mid C_1,C_0)$ and a linearizeable model for $\log\{f(M\mid C_1,a,C_0)/f(M\mid C_1,a',C_0)\}$.

It follows that $\hat{\beta}_a$ and $\hat{\beta}_b$ will be consistent only if their corresponding plugged-in nuisance parameter estimates are consistently estimated. Specifically, $\hat{\beta}_a$ is consistent under correctly specified working parametric submodels $f^W(A\mid C_0;\gamma_1)$ and $M^{\rm ratio;W}(\gamma_2)$, with the remainder of the likelihood left unrestricted; $\hat{\beta}_b$ is consistent under correctly specified working parametric submodels $f^W(A\mid C_0;\gamma_1)$, $C_1^{\rm ratio;W}(\gamma_3)$, and $E^W(Y\mid M,C_1,A,C_0;\gamma_4)$, with the remainder of the likelihood left unrestricted.

The estimator $\hat{\beta}_a$ is an inverse probability of treatment weighted-like estimator, and can be useful if the analyst prefers to leave the conditional distributions of $Y$ and $C_1$ unrestricted. The estimator $\hat{\beta}_b$ can be useful if the analyst prefers to leave the conditional distribution of $M$ unrestricted.

\subsection{Locally efficient estimator}
We now propose a locally semiparametric efficient estimator, $\hat{\beta}_{mr}$, that we will refer to as the multiply robust estimator for reasons we will explain. This estimator is derived from an estimating equation involving the efficient influence function of $\beta_0$ in $\mathcal{M}_{np}$. The efficient influence function is an extension of the parametric score function to semiparametric and nonparametric models. An asymptotically linear estimator with influence function equal to the efficient influence function in a submodel of $\mathcal{M}_{np}$ achieves the minimum asymptotic variance of all regular, asymptotically linear estimators in $\mathcal{M}_{np}$, and is said to be locally semiparametric efficient \citep{bickel1998efficient}.

We present the efficient influence function of $\mathcal{M}_{np}$ in the following theorem. Define $B( \allowbreak m, \allowbreak c_1, \allowbreak a', \allowbreak c_0) \allowbreak  \equiv \allowbreak   E( \allowbreak Y\mid  \allowbreak m, \allowbreak c_1, \allowbreak a', \allowbreak c_0)$, $B' \allowbreak ( \allowbreak c_1, \allowbreak a', \allowbreak a, \allowbreak c_0) \allowbreak  \equiv \allowbreak   E\{ \allowbreak  E( \allowbreak Y\mid  \allowbreak M, \allowbreak c_1, \allowbreak a', \allowbreak c_0)\mid  \allowbreak c_1, \allowbreak a, \allowbreak c_0\}$, and $B''( \allowbreak a', \allowbreak a, \allowbreak c_0) \allowbreak  \equiv \allowbreak   E[ \allowbreak  E\{ \allowbreak  E( \allowbreak Y\mid  \allowbreak M, \allowbreak C_1, \allowbreak a', \allowbreak c_0)\mid  \allowbreak C_1, \allowbreak a, \allowbreak c_0\}\mid  \allowbreak a', \allowbreak c_0]$.

\begin{theorem}
The efficient influence function of $\beta_0$ in $\mathcal{M}_{np}$ is
\begin{align*}
{\rm EIF}(\beta_0) =&\frac{1_{a'}(A)}{f(a'\mid C_0)}M^{\rm ratio}(\allowbreak M,\allowbreak C_1,\allowbreak C_0)\left\{Y - B(M,C_1,a',C_0)\right\}\\
&+ \frac{1_a(A)}{f(a\mid C_0)}\left\{C_1^{\rm ratio}(\allowbreak C_1,\allowbreak C_0)\right\}^{-1}\left\{B(M,C_1,a',C_0) - B'(C_1,a',a,C_0)\right\}\\
&+ \frac{1_{a'}(A)}{f(a'\mid C_0)}\left\{B'(C_1,a',a,C_0) - B''(a',a,C_0)\right\} + \left\{B''(a',a,C_0) - \beta_0\right.\},
\end{align*}
and the asymptotic variance of any regular, asymptotically linear estimator of $\beta_0$ in $\mathcal{M}_{np}$ can be no smaller than $\mathrm{var}\{{\rm EIF}(\beta_0)\}$, the semiparametric efficiency bound for $\mathcal{M}_{np}$.
\end{theorem}
The multiply robust estimator is the M-estimator (or Z-estimator \citep{van2000asymptotic}) solving the estimating equation formed by setting the empirical mean of the estimated efficient influence function to zero for $\beta_0$. The estimator is then
\begin{align*}
\hat{\beta}_{mr} = \mathbb{P}_n\biggl[&\frac{1_{a'}(A)}{\hat{f}(a'\mid C_0)}\hat{M}^{\rm ratio}(\allowbreak M,\allowbreak C_1,\allowbreak C_0)\left\{Y - \hat{B}(M,C_1,a',C_0)\right\}\biggr.\\
+& \frac{1_a(A)}{\hat{f}(a\mid C_0)}\left\{\hat{C}_1^{\rm ratio}(\allowbreak C_1,\allowbreak C_0)\right\}^{-1}\left\{\hat{B}(M,C_1,a',C_0) - \hat{B}'(C_1,a',a,C_0)\right\}\\
+& \biggl.\frac{1_{a'}(A)}{\hat{f}(a'\mid C_0)}\left\{\hat{B}'(C_1,a',a,C_0) - \hat{B}''(a',a,C_0)\right\} + \hat{B}''(a',a,C_0)\biggr].
\end{align*}
It is a function of $f_{M\mid C_1,A,C_0}$ and $f_{C_1\mid A,C_0}$ only via the conditional density ratios $M^{\rm ratio}$ and $C_1^{\rm ratio}$ and conditional expectation functions $B'$ and $B''$.

All nuisance functions are estimated using low-dimensional parametric working models, which we parametrize with $\gamma=(\gamma_1,\gamma_2,\gamma_3,\gamma_2,\gamma_3,\gamma_4,\gamma_5,\gamma_6)$. In particular, $B$ is estimated under the working model $B^W(\gamma_1)$, $B'$ under $B'^W(\gamma_2\mid \gamma_1)= E^W\{B^W(\gamma_1)\mid M,C_1,a',C_0;\gamma_2\}$, and $B''$ under $B''^W(\gamma_3\mid \gamma_1,\gamma_2)= E\{B'^W(\gamma_1,\gamma_2)\mid C_1,a,C_0;\gamma_3\}$. For model robustness purposes, the conditional expectation function $B'$ only requires correct specification in terms of the function of $M$ based on the working model for $Y$, rather than in terms of the true function $B(M,C_1,a',C_0)$, so that $B'^W(\gamma_2\mid \gamma_1)$ can be correctly specified regardless of whether $B^W(\gamma_1)$ is. Likewise, $B''$ only requires correct specification of the function of $C_1$ based on the working models for $M$ and $Y$, so that $B''^W(\gamma_3\mid \gamma_1,\gamma_2)$ can be correctly specified regardless of whether $B^W(\gamma_1)$ and $B'^W(\gamma_2\mid \gamma_1)$ are.

Additionally, $f_{A\mid C_0}$ is estimated under the working model $f^W_{A\mid C_0}(\gamma_4)$, $C_1^{\rm ratio}$ under $C_1^{\rm ratio;W}(\allowbreak \gamma_5)$, and $M^{\rm ratio}$ under $M^{\rm ratio;W}(\gamma_6)$. The latter two models can be formulated as in Section 4.1 using Bayes' theorem and the same procedure to ensure model compatibility. To ensure the models for $B'$ and $B''$ are compatible with these density ratio models, one can use regression models that agree with the models used to obtain $f(a\mid C_1,C_0)$ and $f(a\mid M,C_1,C_0)$ in the procedure described in Section 4.1. For instance, if $B^W(\gamma_1)$ is linear in $M$, and one uses the model $M\mid C_1,A,C_0\sim N(\eta_0+\eta_1C_0+\eta_2A+\eta_3C_1,\sigma^2)$ to obtain a model for $C_1^{\rm ratio}$, then one should use a regression model for $B'$ that is linear in $C_0$, $A$, and $C_1$.

An attractive property of $\hat{\beta}_{mr}$ is its robustness to multiple types of model mis-specification. Using the notation $\theta_M \equiv \left\{B', M^{\rm ratio}\right\}$, $\theta_{C_1} \equiv \left\{B'', C_1^{\rm ratio}\right\}$, $\theta_M^W(\allowbreak \gamma_2,\allowbreak \gamma_6\mid \allowbreak \gamma_1) \allowbreak \equiv \allowbreak \{B'^W(\allowbreak \gamma_2\mid\allowbreak \gamma_1),\allowbreak  M^{\rm ratio;W}(\allowbreak \gamma_6)\}$, and $\theta^W_{C_1}(\gamma_3,\gamma_5\mid\gamma_1,\gamma_2) \equiv \left\{B''^W(\gamma_3\mid\gamma_1,\gamma_2), C_1^{\rm ratio;W}(\gamma_5)\right\}$, we give a multiple robustness result in the following theorem.
\begin{theorem}
The estimator $\hat{\beta}_{mr}$ is consistent and asymptotically normal (under standard regularity conditions) provided that one of the following holds: (a) $\{\theta_M,f_{A\mid C_0}\} \in \{\theta_M^W(\gamma_2,\gamma_6\mid\gamma_1),f_{A\mid C_0}^W(\gamma_4)\}$, (b) $\{B,\theta_{C_1},f_{A\mid C_0}\} \in \{B^W(\gamma_1),\theta_{C_1}^W(\gamma_3,\gamma_5\mid\gamma_1,\gamma_2),f_{A\mid C_0}^W(\gamma_4)\}$, (c) $\{B, \allowbreak \theta_{C_1}, \allowbreak \theta_M\}\allowbreak  \in \allowbreak \{B^W(\gamma_1), \theta_{C_1}^W(\gamma_3,\gamma_5\mid\gamma_1,\gamma_2),\theta_M^W(\gamma_2,\gamma_6\mid\gamma_1)\}$. It is locally semiparametric efficient in that it will achieve the semiparametric efficiency bound in the intersection submodel in which (a)--(c) all hold.
\end{theorem}
Thus, $\hat{\beta}_{mr}$ offers three distinct opportunities to obtain a consistent estimator of the path-specific effect. By contrast, $\hat{\beta}_a$ will be consistent only if a slightly weaker form of (a) holds, where $B'^W(\gamma_2)$ need not be correctly specified; $\hat{\beta}_b$ will be consistent only if a slightly weaker form of (b) holds, where $B''^W(\gamma_3)$ need not be correctly specified; and $\hat{\beta}_{mle}$ will be consistent only if a slightly weaker form of (c) holds, where $M^{\rm ratio;W}(\gamma_6)$ and $C_1^{\rm ratio;W}(\gamma_5)$ need not be correctly specified.

For inference on $\hat{\beta}_{mr}$, we recommend the nonparametric bootstrap \citep{efron1979bootstrap} or similar alternative resampling methods. While one might consider using an empirical estimator of the efficient influence function variance, it does not have the multiple-robustness property, since $\hat{\beta}_{mr}$ is not globally efficient. Thus, this variance estimator may be inconsistent under certain forms of model mis-specification even if $\hat{\beta}_{mr}$ is still consistent.

While all four estimators given here are in fact asymptotically equivalent under a nonparametric model, they will have different asymptotic properties under parametric and semiparametric models \citep{tchetgen2012semiparametric}.

\subsection{Stabilization techniques}
Due to the proposed estimators' reliance on inverse-probability weights, they may suffer from instability in settings where the set of positivity assumptions is nearly violated \citep{kang2007demystifying}. A useful stabilization technique is to simply replace any propensity score $\hat{f}_{A\mid X}$ with $\hat{f}^{\dagger}_{A\mid  X}$, where $X$ is some vector of covariates and $\mathrm{logit} \allowbreak \hat{f}^{\dag}_{A\mid  X}( \allowbreak a\mid  \allowbreak  X) \allowbreak = \allowbreak \mathrm{logit} \allowbreak \hat{f}_{A\mid  X}( \allowbreak a\mid  \allowbreak  X) \allowbreak - \allowbreak \log[ \allowbreak \mathbb{P}_n\{ \allowbreak 1_{a'}(A)\}] \allowbreak + \allowbreak \log[ \allowbreak \mathbb{P}_n\{ \allowbreak 1_a(A) \allowbreak \hat{f}_{A\mid  X}( \allowbreak a'\mid  \allowbreak  X) \allowbreak / \allowbreak \hat{f}_{A\mid  X}( \allowbreak a\mid  \allowbreak  X)\}],$ which ensures the weights are bounded, as discussed in \cite{tchetgen2012semiparametric}.

Another stabilization technique, based on a procedure proposed by \cite{robins2000robust} and detailed in his comment \citep{robins2007comment} in response to \cite{kang2007demystifying}, can also be adapted to this setting for the multiply robust estimator. The idea is to obtain a substitution estimator by carefully selecting regression models and an estimation strategy such that the three terms in $\hat{\beta}_{mr}$ depending on weights are empirically evaluated as null, leaving the term $\mathbb{P}_n\hat{B}''(a',a,C_0;\hat{\gamma}_3\mid\hat{\gamma}_1,\hat{\gamma}_2)$, a plug-in term, which does not depend on weights. This can be accomplished by the following procedure.

First, fit propensity score models to obtain estimates $\hat{f}_{A\mid C_0}(\hat{\gamma}_4)$, $\hat{C}_1^{\rm ratio}(\hat{\gamma}_5)$, and $\hat{M}^{\rm ratio}(\hat{\gamma}_6)$ as described previously. Next, using these estimates, estimate $\gamma_1$ by solving
\begin{align*}
\mathbb{P}_n\Biggl[&\frac{1_{a'}(A)}{\hat{f}(a'\mid C_0;\hat{\gamma}_4)}\hat{M}^{\rm ratio}(\allowbreak M,\allowbreak C_1,\allowbreak C_0;\hat{\gamma}_6)\Biggr.\\
\times &\Biggl.\left\{\nabla_{\gamma_1}B(M,C_1,a',C_0;\gamma_1)\right\}\left\{Y - B(M,C_1,a',C_0;\gamma_1)\right\}\Biggr]=0,
\end{align*}
where $B^W(\gamma_1)$ contains an intercept, such that one of the elements in $\nabla_{\gamma_1}B(M,C_1,a',C_0;\gamma_1)$ is one. This ensures that the first term in the estimating equation for $\hat{\beta}_{mr}$ is zero at $\hat{\gamma}$. Next, estimate $\gamma_2$ by solving
\begin{align*}
\mathbb{P}_n\Biggl[\frac{1_a(A)}{\hat{f}(a\mid C_0;\hat{\gamma}_4)}\left\{\hat{C}_1^{\rm ratio}(\allowbreak C_1,\allowbreak C_0;\hat{\gamma}_5)\right\}^{-1}\left\{\nabla_{\gamma_2}B'(C_1,a',C_0;\gamma_2\mid\hat{\gamma}_1)\right\}\Biggr.&\\
\times \Biggl.\left\{\hat{B}(M,C_1,a',C_0;\hat{\gamma}_1) - B'(C_1,a',a,C_0;\gamma_2\mid\hat{\gamma}_1)\right\}&\Biggr]=0,
\end{align*}
where $B'^W(\gamma_2\mid\hat{\gamma}_1)$ contains an intercept, such that one of the elements in $\nabla_{\gamma_2}B'(C_1,a',C_0;\gamma_2\mid\hat{\gamma}_1)$ is one. This ensures that the second term in the estimating equation for $\hat{\beta}_{mr}$ is zero at $\hat{\gamma}$. Next, estimate $\gamma_3$ by solving
\begin{align*}
\mathbb{P}_n\Biggl[&\frac{1_{a'}(A)}{\hat{f}(a'\mid C_0;\hat{\gamma}_4)}\left\{\nabla_{\gamma_3}B''(a',C_0;\gamma_3\mid\hat{\gamma}_1,\hat{\gamma}_2)\right\}\Biggr.\\
\times &\Biggl.\left\{\hat{B}'(C_1,a',a,C_0;\hat{\gamma}_2\mid\hat{\gamma}_1) - \hat{B}''(a',a,C_0;\gamma_3\mid\hat{\gamma}_1,\hat{\gamma}_2)\right\}\Biggr]=0,
\end{align*}
where $B''^W(\gamma_3\mid\hat{\gamma}_1,\hat{\gamma}_2)$ contains an intercept, such that one of the elements in $\nabla_{\gamma_3}B''(a',C_0;\gamma_3\mid\hat{\gamma}_1,\hat{\gamma}_2)$ is one. This ensures that the third term in the estimating equation for $\hat{\beta}_{mr}$ is zero at $\hat{\gamma}$. Finally, plugging $\hat{\gamma}_1$, $\hat{\gamma}_2$, and $\hat{\gamma}_3$ into $\hat{\beta}_{mr}$ leaves $\mathbb{P}_n\hat{B}''(a',a,C_0;\hat{\gamma}_3\mid\hat{\gamma}_1,\hat{\gamma}_2)$, as desired. If $B^W(\gamma_1)$, $B'^W(\gamma_2\mid\hat{\gamma}_1)$, and $B''^W(\gamma_3\mid\hat{\gamma}_1,\hat{\gamma}_2)$ are all linear models, then this procedure can be accomplished by iteratively fitting linear models with intercepts for $B$, $B'$, and $B''$ using weighted least squares weights with weights equal to $1_{a'}(A)/\hat{f}(a'\mid C_0;\hat{\gamma}_4)\times\hat{M}^{\rm ratio}(\allowbreak M,\allowbreak C_1,\allowbreak C_0;\hat{\gamma}_6)$, $1_a(A)/\{\hat{f}(a\mid C_0;\hat{\gamma}_4)\hat{C}_1^{\rm ratio}(\allowbreak C_1,\allowbreak C_0;\hat{\gamma}_5)\}$, and $1_{a'}(A)/\hat{f}(a'\mid C_0;\hat{\gamma}_1)$, respectively.

This latter stabilized estimator matches the targeted minimum-loss based estimator \citep{van2006targeted} for a particular choice of submodels and loss functions. For example, if linear models are used for $B$, $B'$, and $B''$, then the estimator described above corresponds to a targeted minimum loss-based estimator using squared-error loss functions for $B$, $B'$, and $B''$, submodels
\begin{align*}
B(\epsilon_1)(C_0,C_1,M) &= B_0(C_0,C_1,M)+\frac{1_{a'}(A)}{f(a'\mid C_0)}M^{\rm ratio}(\allowbreak M,\allowbreak C_1,\allowbreak C_0)(1,C_0,C_1,M)\epsilon_1\\
B'(\epsilon_2)(C_0,C_1) &= B'_0(C_0,C_1)+\frac{1_a(A)}{f(a\mid C_0)}\left\{C_1^{\rm ratio}(\allowbreak C_1,\allowbreak C_0)\right\}^{-1}(1,C_0,C_1)\epsilon_2\\
B''(\epsilon_3)(C_0) &= B''_0(C_0)+\frac{1_{a'}(A)}{f(a'\mid C_0)}(1,C_0)\epsilon_3,\\
\end{align*}
and using initial estimates $B_0(C_0,C_1,M) = B'_0(C_0,C_1) = B''_0(C_0) = 0$.

\section{Simulation study}
We conducted a simulation study in order to demonstrate the finite-sample performance of these estimators as well as the multiple-robustness property of $\hat{\beta}_{mr}$. We generated 1000 data sets of size 5000 from the data generating mechanism:
\begin{align*}
C_0 &\sim \mathcal{U}(0,2)\\
A\mid C_0 &\sim Bernoulli\left[1-\{1+\exp(0.9+0.3C_0)\}^{-1}\right]\\
C_1 &= \begin{pmatrix} 0.8\\ 0.6\\ -0.3 \end{pmatrix} + \begin{pmatrix} 1\\ 0.1\\ 0.2 \end{pmatrix} C_0 + \begin{pmatrix} 0.5\\ -0.4\\ 0.5 \end{pmatrix} A + \begin{pmatrix} -0.1\\ 0.8\\ -0.2 \end{pmatrix} C_0 A + \mathcal{N}(0,I)\\
M &= -0.5 - 0.2C_0 + 0.3A + [-0.2, 0.1, 0.5]C_1 + [0.4, 0, 0]AC_1 + N(0,1)\\
Y &= 0.2 + 0.2C_0 + 0.6A + [1, 0.7, 0.3]C_1 - 0.9M - 0.8AM + N(0,1).
\end{align*}
In order to investigate the impact of model mis-specification, we computed each of the four estimators given above, $\hat{\beta}_{mr}$, $\hat{\beta}_{mle}$ $\hat{\beta}_a$, and $\hat{\beta}_b$, under the four parametric models, $\mathcal{M}_a$, $\mathcal{M}_b$, $\mathcal{M}_c$, and $\mathcal{M}_{int}$. Models $\mathcal{M}_a$, $\mathcal{M}_b$, and $\mathcal{M}_c$ were specified such that statements (a)--(c) in Section 4.2 corresponding to their respective subscripts held, but the models for the remaining estimands were incorrectly specified. For instance, under $\mathcal{M}_a$, models $\theta_M^W$ and $f_{A\mid C_0}^W$ are correctly specified, while $B^W$ and $\theta_{C_1}^W$ are not. The intersection model uses correctly specified working models. All models were fit by maximum likelihood. The first stabilization technique described in the previous section was used to adjust propensity scores. We used the following working models, subscripted $C$ for correctly specified and $I$ for incorrectly specified, and where $\Phi$ denotes the standard normal distribution function:

\noindent $f_{A\mid C_0}^{W}$:\\
\hspace*{2em} Correct: $\textrm{logit pr}_C\{A = 1\mid C_0\} = [1, C_0]\alpha_C$\\
\hspace*{2em} Incorrect: $\Phi^{-1}(\textrm{pr}_I\{A = 1\mid C_0\}) = [1, C_0]\alpha_I$

\noindent $B^{W}$:\\
\hspace*{2em} Correct: $ E_C[Y\mid M,C_1,A,C_0] = [1, C_0, A, C_1, M, AM]\eta_C$\\
\hspace*{2em} Incorrect: $ E_I[Y\mid M,C_1,A,C_0] = [1, C_0, A, C_1, M]\eta_I$

\noindent $\theta_{C_1}^W$:\\
\hspace*{2em} Correct: $C_1^{\rm ratio;W} = \textrm{pr}_C(A=a\mid C_1,C_0)/\textrm{pr}_C( \allowbreak A \allowbreak = \allowbreak a'\mid  \allowbreak C_1, \allowbreak C_0) \allowbreak \times \allowbreak  \textrm{pr}_C( \allowbreak A \allowbreak = \allowbreak a'\mid  \allowbreak C_0) / \allowbreak \textrm{pr}_C( \allowbreak A \allowbreak = \allowbreak a\mid  \allowbreak C_0)$,
which depends on the correctly specified $f_{A\mid C_0}^{W}$ model and the correctly specified model
$\textrm{logit pr}_C\{A = 1\mid C_1,C_0\} = [1, C_0, C_0^2, C_1, C_0C_1]\lambda_C$;\\
$B''_C(a',a,C_0)= E_C[ E_C\{ E_C(Y\mid M,C_1,a',C_0)\mid C_1, \allowbreak a, \allowbreak C_0\}\mid  \allowbreak a', \allowbreak C_0]$,
which depends on the correctly specified $B^{W}$ model and the correctly specified models $ E_C[ \allowbreak C_{1j}\mid  \allowbreak A, \allowbreak C_0] \allowbreak  = \allowbreak  [1, \allowbreak  C_0, \allowbreak  A, \allowbreak  C_0A] \allowbreak \delta_{j;C}$ for all $j \allowbreak \in \allowbreak  \{1, \allowbreak 2, \allowbreak 3\}$ and $ E_C[ \allowbreak M\mid  \allowbreak C_1, \allowbreak A, \allowbreak C_0] \allowbreak  = \allowbreak  [1, \allowbreak  C_0, \allowbreak  A, \allowbreak  C_1, \allowbreak  AC_{11}] \allowbreak \zeta_C$.\\
\hspace*{2em} Incorrect: $C_1^{\rm ratio;W,I} = \textrm{pr}_I(A=a\mid C_1,C_0)/\textrm{pr}_I(A=a'\mid C_1,C_0)\times\textrm{pr}_C(A=a'\mid C_0)/\textrm{pr}_C(A=a\mid C_0)$, which depends on the correctly specified $f_{A\mid C_0}^{W}$ model and the incorrectly specified model
$\textrm{logit pr}_I\{A = 1\mid C_1,C_0\} = [1, C_0, C_1]\lambda_I$;\\
$B''_I(a',a,C_0)= E_I[ E_C\{ E_I(Y\mid M,C_1,a',C_0)\mid C_1,a,C_0\}\mid a',C_0]$,
which depends on the incorrectly specified $B^{W}$ model, the correctly specified working mean model for $M$ used for $B''_C(a',a,C_0)$ above, and the incorrectly specified model $ E_I[C_{1j}\mid A,C_0] = [1, C_0, A]\delta_{j,I}$,
since $\theta_{C_1}^W$ is only mis-specified in setting (a), under which $B^W$ is also mis-specified and $\theta_M^W$ is correctly specified.

\noindent $\theta_M^W$:\\
\hspace*{2em} Correct: $M^{\rm ratio;W,C} = \textrm{pr}_C(A=a\mid M,C_1,C_0) \allowbreak / \allowbreak \textrm{pr}_C( \allowbreak A \allowbreak = \allowbreak a'\mid  \allowbreak M, \allowbreak C_1, \allowbreak C_0) \allowbreak \times \allowbreak \textrm{pr}_C( \allowbreak A \allowbreak = \allowbreak a'\mid  \allowbreak C_1, \allowbreak C_0) \allowbreak / \allowbreak \textrm{pr}_C( \allowbreak A \allowbreak = \allowbreak a\mid  \allowbreak C_1, \allowbreak C_0)$,
which depends on the correctly specified model $\textrm{logit pr}_C \allowbreak \{A \allowbreak  = \allowbreak  1\mid  \allowbreak M, \allowbreak C_1, \allowbreak C_0\}  \allowbreak = \allowbreak  [1, \allowbreak  C_0, \allowbreak  C_0^2, \allowbreak  C_1, \allowbreak  C_0C_1, \allowbreak  C_{11}C_1, \allowbreak  M, \allowbreak  C_{11}M] \allowbreak \gamma_C$
and the correctly specified logistic model used for $C_1^{\rm ratio;W}$ above;\\
$B'_C( \allowbreak C_1, \allowbreak a', \allowbreak a, \allowbreak C_0) \allowbreak  = \allowbreak   E_C\{ \allowbreak  E_C( \allowbreak Y\mid  \allowbreak M, \allowbreak C_1, \allowbreak a', \allowbreak C_0)\mid  \allowbreak C_1, \allowbreak a, \allowbreak C_0\}$ depends on the correctly specified $B^{W}$ model and the correctly specified mean model for $M$ used for $B''_C(a', \allowbreak a, \allowbreak C_0)$ above.

Incorrect: $M^{\rm ratio;W,I} = \textrm{pr}_I(A=a\mid M,C_1,C_0) \allowbreak / \allowbreak \textrm{pr}_I( \allowbreak A \allowbreak = \allowbreak a'\mid  \allowbreak M, \allowbreak C_1, \allowbreak C_0) \allowbreak \times \allowbreak \textrm{pr}_C( \allowbreak A \allowbreak = \allowbreak a'\mid  \allowbreak C_1, \allowbreak C_0) \allowbreak / \allowbreak \textrm{pr}_C( \allowbreak A \allowbreak = \allowbreak a\mid  \allowbreak C_1, \allowbreak C_0)$, which depends on the correctly specified logistic model for $\textrm{pr}_C\{ \allowbreak A \allowbreak  = \allowbreak  1\mid  \allowbreak C_1, \allowbreak C_0\}$ and the incorrectly specified model
$\textrm{logit pr}_I\{ \allowbreak A \allowbreak  = \allowbreak  1\mid  \allowbreak M, \allowbreak C_1, \allowbreak C_0\} \allowbreak  = \allowbreak  [1, \allowbreak  C_0, \allowbreak  C_1, \allowbreak  M] \allowbreak \gamma_I$;\\
$B'_I( \allowbreak C_1, \allowbreak a', \allowbreak a, \allowbreak C_0) \allowbreak  = \allowbreak   E_I\{ \allowbreak  E_C( \allowbreak Y\mid  \allowbreak M, \allowbreak C_1, \allowbreak a', \allowbreak C_0)\mid  \allowbreak C_1, \allowbreak a, \allowbreak C_0\}$,
which depends on the incorrectly specified model $ E_I[ \allowbreak M\mid  \allowbreak C_1, \allowbreak A, \allowbreak C_0] \allowbreak  = \allowbreak  [1, \allowbreak  C_0, \allowbreak  A, \allowbreak  C_1] \allowbreak \zeta_I$ and the correctly specified model $B^{W,C}$, since $\theta_M^W$ is only mis-specified in setting (c), under which $B^W$ is correctly specified.

\begin{figure}
\centering
\includegraphics[scale=.4]{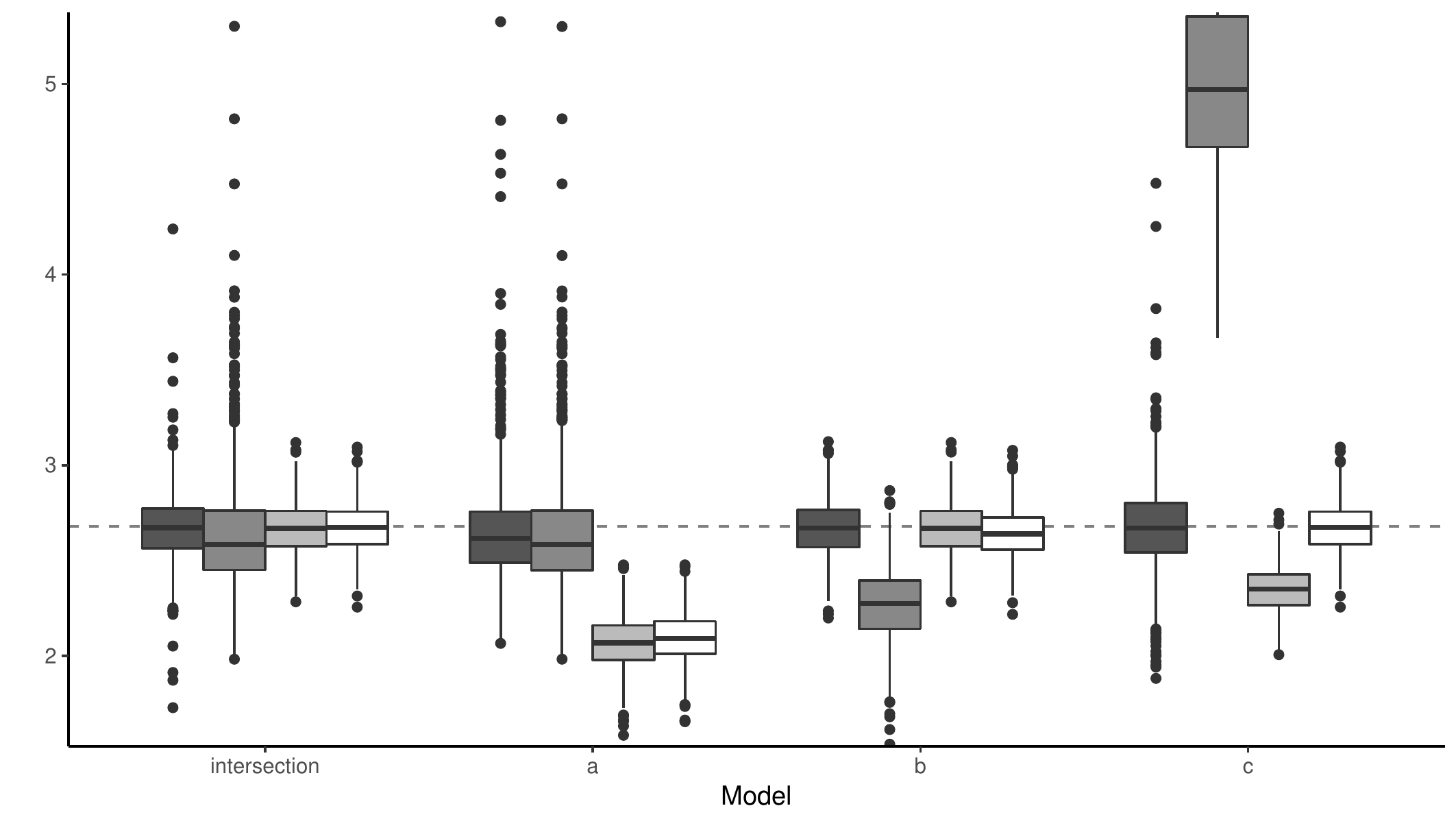}
\caption{Simulation results for n=5000. Boxplots of each of the four $\mathcal{P}_{AMY}$-specific estimators are given under $\mathcal{M}_{int}$, $\mathcal{M}_a$, $\mathcal{M}_b$, and $\mathcal{M}_c$. The dark gray boxplot represents the multiply robust etimator; the medium gray boxplot represents estimator a, the light gray boxplot represents estimator b; the white boxplot represents the maximum likelihood estimator; the gray horizontal dashed line is through the true parameter value, $\beta_0$.\label{fig:sims}}
\end{figure}

The results are summarized in the boxplots of the four estimators displayed in Fig.~\ref{fig:sims}. 
All estimators are roughly centered around $\beta_0$ under $\mathcal{M}_{int}$. Besides $\hat{\beta}_{mr}$, $\hat{\beta}_a$ is the only consistent estimator under $\mathcal{M}_a$, $\hat{\beta}_b$ is the only consistent estimator under $\mathcal{M}_b$, and $\hat{\beta}_{mle}$ is the only consistent estimator under $\mathcal{M}_c$. The estimator $\hat{\beta}_{mr}$ is consistent under all models. Therefore, in moderate to large samples, we expect to see the estimators consistent under each model to concentrate around $\beta_0$ accordingly.

The results plainly illustrate the multiple-robustness property of $\hat{\beta}_{mr}$. As predicted, while the other estimators failed to estimate $\beta_0$ without bias, the $\hat{\beta}_{mr}$ concentrated around $\beta_0$ under every model. For the other estimators, each concentrated around $\beta_0$ under $\mathcal{M}_{int}$ and their corresponding models where the mis-specified components did not factor into estimation, as expected. That is, $\hat{\beta}_a$ concentrated around $\beta_0$ under $\mathcal{M}_a$, $\hat{\beta}_b$ concentrated around $\beta_0$ under $\mathcal{M}_b$, and $\hat{\beta}_{mle}$ concentrated around $\beta_0$ under $\mathcal{M}_c$. The estimators did exhibit noticeable bias, however, under the other models, with the exception of the minimal bias exhibited by $\hat{\beta}_c$ under $\mathcal{M}_b$. Thus, all estimators other than $\hat{\beta}_{mr}$ were biased under at least one model. We see a trade-off between efficiency and robustness; in all settings, $\hat{\beta}_{mle}$ and $\hat{\beta}_b$ perform best in terms of efficiency, with a slight advantage going to $\hat{\beta}_{mle}$, as expected. While $\hat{\beta}_{mr}$ and $\hat{\beta}_a$ roughly concentrate around $\beta_0$ under $\mathcal{M}_a$, they are fairly right skewed, indicating that their asymptotic distributions have not yet come into focus.

\section{Harvard PEPFAR Nigeria analysis}
We now present results of the Harvard PEPFAR data analysis. The data set consisted of 48,345 observations, 9968 (41.9\%) of which were complete observations, i.e., observations with no missing variables. Nonmonotone missingness was handled by multivariate imputation by chained equations \citep{buuren2011mice}.

Our effect of interest is the path-specific effect of treatment regimen assignment on log CD4 count through adherence on the mean difference scale. We estimated $\beta_0$ with $\hat{\beta}_{mr}$ and $ E\{Y(a')\}$ with the augmented inverse probability of treatment weighted estimator \citep{bang2005doubly}. Comparisons with other estimators are available in the supplementary materials. Results were fairly consistent across estimators. Let $\hat{\mathcal{P}}_{AMY;mr}$ denote the effect estimate contrasting these two estimators. We computed this estimate and its corresponding confidence interval using a bootstrap variance estimate for each pairwise comparison of treatments.

We coded the treatment regimens in descending order of magnitude of their total effects on mean log CD4 count. That is, they were coded in ascending order of mean counterfactual CD4 count had everyone been assigned to that treatment, since a lower counterfactual risk of failure corresponds to a higher magnitude of total effect. See Table \ref{tab:cd4} note for treatment coding. The order of these effects differed from those on risk of virological failure, so our coding does not correspond directly with that used in \cite{doi:10.1080/01621459.2017.1295862}. Because in practice we are more interested in learning how less-effective treatments can be improved, we only consider the higher-coded treatment in a pair as the baseline, $a'$.

We are primarily interested in the percentage of the total effect attributable to the mediated effect, i.e., the percent mediated by $\mathcal{P}_{AMY}$. If this value is close to 100\%, we can conclude that the drugs themselves likely have the same effectiveness on CD4 count, and that it is their differential effect on adherence not due to toxicity that is driving the difference in total effects. If, on the other hand, this percentage is small or negative, we can only say that the difference in total effects is not driven by a difference in effects through $\mathcal{P}_{AMY}$. It may be the case that the efficacies of the drugs themselves do, in fact, differ, or that the difference in total effects is driven by the differential effect on adherence due to toxicity, but we cannot confirm either. Table \ref{tab:cd4} shows one hundred times $\hat{\mathcal{P}}_{AMY;mr}$ divided by the total effect estimates, which are also on the mean difference scale and are estimated with the augmented inverse probability of treatment weighted estimator.
\begin{table}[h]
\begin{center}
\caption{Estimated percentage of total effect on log CD4 count due to $\mathcal{P}_{AMY}$-specific effect\label{tab:cd4}}
\begin{tabular}{ c  r@{}l r@{}l r@{}l r@{}l}
\\
&\multicolumn{8}{c}{Baseline treatment}\\
Comparison trt &\multicolumn{2}{c}{2}&\multicolumn{2}{c}{3}&\multicolumn{2}{c}{4}&\multicolumn{2}{c}{5}\\
1 & -2& & 44&$^*$ & 7& & -3	&$^*$ \\
2 & \multicolumn{2}{c}{-} & -103	& & 9	& & -4& \\
3 & \multicolumn{2}{c}{-} & \multicolumn{2}{c}{-} & 4	& & -11&$^*$ \\
4 & \multicolumn{2}{c}{-} & \multicolumn{2}{c}{-} & \multicolumn{2}{c}{-} & -54&$^*$ \\
\\
\end{tabular}
\end{center}
NOTE: $^*$Significant path-specific effect ($\alpha=0.05$). 1 = AZT + 3TC + NVP, 2 = TDF + 3TC/FTC + EFV, 3 = AZT + 3TC + EFV, 4 = d4T + 3TC + NVP, 5 = TDF + 3TC/FTC + NVP. 3TC=lamivudine, AZT=zidovudine, d4T=stavudine, EFV=efavirenz, FTC=emtricitabine, NVP=nevirapine, TDF=tenofovir.
\end{table}
\begin{comment}
\begin{table}[h]
\begin{center}
\caption{Estimated percentage of total effect on virologic failure due to $\mathcal{P}_{AMY}$-specific effect\label{tab:vf}}
\begin{tabular}{ c r@{}l r@{}l r@{}l r@{}l}
\\
&\multicolumn{8}{c}{Baseline treatment}\\
Comparison trt &\multicolumn{2}{c}{2}&\multicolumn{2}{c}{3}&\multicolumn{2}{c}{4}&\multicolumn{2}{c}{5}\\
1 & 13& & 0& & 7& & -3& \\
2 & \multicolumn{2}{c}{-} & -15&$^*$ & -2& & -13&$^*$ \\
3 & \multicolumn{2}{c}{-} & \multicolumn{2}{c}{-} & 4& & -10&$^*$ \\
4 & \multicolumn{2}{c}{-} & \multicolumn{2}{c}{-} & \multicolumn{2}{c}{-} & -59&$^*$ \\
\\
\end{tabular}
\end{center}
NOTE: $^*$Significant path-specific effect ($\alpha=0.05$).
\end{table}
\begin{table}[h]
\begin{center}
\caption{Estimated proportion of total effect on virologic failure due to $\mathcal{P}_{AMY}$-specific effect\label{tab:vf}}
\begin{tabular}{ c r@{.}l r@{.}l r@{.}l r@{.}l}
\\
&\multicolumn{8}{c}{Baseline treatment}\\
Comparison trt &\multicolumn{2}{c}{2}&\multicolumn{2}{c}{3}&\multicolumn{2}{c}{4}&\multicolumn{2}{c}{5}\\
1 & 0&13 & 0&001 & 0&073 & -0&029 \\
2 & \multicolumn{2}{c}{-} & -0&15$^*$ & -0&024 & -0&13$^*$ \\
3 & \multicolumn{2}{c}{-} & \multicolumn{2}{c}{-} & 0&036 & -0&10$^*$ \\
4 & \multicolumn{2}{c}{-} & \multicolumn{2}{c}{-} & \multicolumn{2}{c}{-} & -0&59$^*$ \\
\\
\end{tabular}
\end{center}
NOTE: $^*$Significant path-specific effect ($\alpha=0.05$).
\end{table}
\end{comment}
Treatment comparisons with statistically significant path-specific effects are indicated by asterisks. Due to the treatment coding, the denominators of the Table \ref{tab:cd4} values are always positive. Thus, a positive path-specific effect will be in the same direction as the total effect, and hence will explain a positive proportion of it.

We estimated a significant path-specific effect in four of the treatment comparisons. In one of these comparisons, the effect estimate was positive; in the other three it was negative. In the latter case, this implies that the percentages of the total effects due to the effects through $\mathcal{P}_{AMY}$ were also negative, since the total effect estimates are positive for each treatment comparison we consider. Thus, in these treatment comparisons, the estimated $\mathcal{P}_{AMY}$ path-specific effects and estimated total effects are in opposite directions. Since the total effect can be decomposed as a sum of multiple path-specific effects including the one through $\mathcal{P}_{AMY}$, this means that the other path-specific effects are, in sum, stronger in the opposite direction than the $\mathcal{P}_{AMY}$ path-specific effect, and overwhelm it to produce a positive total effect. This also means that had there been no effect through $\mathcal{P}_{AMY}$, the total effect would have been even larger than what we observed, since we would have been adding zero to the other path-specific effects that compose the total effect, rather than a negative value. For example, our findings indicate that the effect of treatment 5, as compared to treatment 4, would have been 54\% lower if its impact on adherence via mechanisms other than toxicity (e.g., pill count, meal restrictions, etc.) were the same as that of treatment 4. This would result in an even larger total effect between these treatments.

The path-specific effect comparing treatment 1 to treatment 3, on the other hand, explains a positive proportion (approximately 44\%) of the total effect estimate. This means that when assigning treatment 3, if we could intervene to change the the factors affecting later adherence other than early adherence and toxicity to be the same as those the patients would experience under treatment 1 (e.g., the same pill count and meal restrictions as treatment 1), then we would be able to close the gap in effectiveness on log CD4 count between treatments 3 and 1 by about 44\%. A final note: the percentage mediated of the effect comparing treatments 3 and 2 is inflated to -103\% by virtue of the  denominator, i.e., the total effect, being fairly small.

\appendix

\section{Theoretical results}

\subsection{Derivation of estimators a and b}
The estimator $\hat{\beta}_a$ arises from an alternative representation of the identifying functional of $\beta_0$:
\begin{align*}
\iiint\limits_{m,c_1,c_0}&\E(Y\mid m,c_1,e',c_0)dF_{M\mid C_1,E,C_0}(m\mid c_1,e,c_0)dF_{C_1\mid E,C_0}(c_1\mid e',c_0)dF_{C_0}(c_0)\\
&= \sum\limits_{e^*\in\{e',e\}}\;\int\limits_{y,m,c_1,c_0}y\frac{1_{e'}(e^*)}{f(e'\mid c_0)}\frac{f(m\mid c_1,e,c_0)}{f(m\mid c_1,e^*,c_0)}dF_{Y,M,C_1,E,C_0}(y,m,c_1,e^*,c_0)\\
&= \E\left\{\frac{1_{e'}(E)}{f(e'\mid C_0)}M^{ratio}Y\right\}.
\end{align*}
We simply plug in the estimates, $\hat{f}_{E=0\mid C_0}$ and $\hat{M}^{ratio}$, and compute the empirical mean. Thus, we have
\[\hat{\beta}_a \equiv \mathbb{P}_n\left\{\frac{1_{e'}(E)}{\hat{f}(e'\mid C_0)}\hat{M}^{ratio}Y\right\}.\]

The estimator $\hat{\beta}_b$ arises from a second representation of the identifying functional of $\beta_0$:
\begin{align*}
\iiint\limits_{m,c_1,c_0}&\E(Y\mid m,c_1,e',c_0)dF_{M\mid C_1,E,C_0}(m\mid c_1,e,c_0)dF_{C_1\mid E,C_0}(c_1\mid e',c_0)dF_{C_0}(c_0)\\
&= \sum\limits_{e^*\in\{e',e\}}\;\int\limits_{m,c_1,c_0}\E(Y\mid M,C_1,e',C_0)\frac{1_e(e^*)}{f(e^*\mid c_0)}\frac{f(c_1\mid e',c_0)}{f(c_1\mid e^*,c_0)}dF_{M,C_1,E,C_0}(m,c_1,e^*,c_0)\\
&= \E\left[\frac{1_e(E)}{f(e\mid C_0)}\left(C_1^{ratio}\right)^{-1}\E(Y\mid M,C_1,e',C_0)\right].
\end{align*}
Again, we plug in the estimates $\hat{C}_1^{ratio}$, $\hat{f}_{E=1\mid C_0}$ and $\hat{\E}(Y\mid M,C_1,e',C_0)$, and compute the empirical mean. Thus, we have
\[\hat{\beta}_b \equiv \mathbb{P}_n\left\{\frac{1_e(E)}{\hat{f}(e\mid C_0)}\left(\hat{C}_1^{ratio}\right)^{-1}\hat{\E}(Y\mid M,C_1,e',C_0)\right\}.
\]

\subsection{Proofs}
\begin{proof}[Proof of Theorem 1]
Let $\nu$ denote the appropriate dominating measure or product measure corresponding to each combination of random variables. Let $F_{O;t}\allowbreak  = F_{Y\mid M,C,E,C_0;t}\allowbreak F_{M\mid C_1,E,C_0;t}\allowbreak F_{C_1\mid E,C_0;t}\allowbreak F_{E\mid C_0;t}\allowbreak F_{C_0;t}$ denote a one-dimensional regular parametric submodel of $\mathcal{M}_{np}$ with $F_{O,0} \allowbreak = \allowbreak F_{O}$, and let
\begin{align*}
\beta_t &= \beta_0(F_{O;t}) = \E_t(Y[M\{e,C_1(e')\},C_1(e'),e'])\\
&= \int\limits_{m,c_1,c_0}\E_t(Y\mid m,c_1,e',c_0)f_t(M=m\mid c_1,e,c_0)f_t(C_1=c_1\mid e',c_0)f_t(C_0=c_0)\\
&\qquad\qquad \times d\nu(m,c_1,c_0)
\end{align*}
and $U_O = \frac{\nabla_{t=0}f_t(O)}{f(O)}$ be the score for $O$. Then
\begin{align*}
&\left.\frac{\partial\beta_t}{\partial t}\right\rvert _{t=0} = \\
&\int\limits_{m,c_1,c_0}\nabla_{t=0}\E_t(Y\mid m,c_1,e',c_0)f(M=m\mid c_1,e,c_0)f(C_1=c_1\mid e',c_0)f(C_0=c_0)\\
&\qquad\qquad\times d\nu(m,c_1,c_0)\tag{1}\\
&+ \int\limits_{m,c_1,c_0}\E(Y\mid m,c_1,e',c_0)\nabla_{t=0}f_t(M=m\mid c_1,e,c_0)f(C_1=c_1\mid e',c_0)f(C_0=c_0)\\
&\qquad\qquad\times d\nu(m,c_1,c_0)\tag{2}\\
&+ \int\limits_{m,c_1,c_0}\E(Y\mid m,c_1,e',c_0)f(M=m\mid c_1,e,c_0)\nabla_{t=0}f_t(C_1=c_1\mid e',c_0)f(C_0=c_0)\\
&\qquad\qquad\times d\nu(m,c_1,c_0)\tag{3}\\
&+ \int\limits_{m,c_1,c_0}\E(Y\mid m,c_1,e',c_0)f(M=m\mid c_1,e,c_0)f(C_1=c_1\mid e',c_0)\nabla_{t=0}f_t(C_0=c_0)\\
&\qquad\qquad\times d\nu(m,c_1,c_0),\tag{4}\\
\end{align*}
where
\begin{align*}
(1) = &\int\limits_{m,c_1,c_0} \nabla_{t=0}\E_t(Y\mid m,c_1,e',c_0)f(M=m\mid c_1,e,c_0)f(C_1=c_1\mid e',c_0)f(C_0=c_0)\\
&\qquad\qquad\times d\nu(m,c_1,c_0)\\
= &\int\limits_{m,c_1,c_0} \int_yy\left\{\frac{\nabla_{t=0}f_t(y,m,c_1,e',c_0)}{f(m,c_1,e',c_0)} - \frac{f(y,m,c_1,e',c_0)\nabla_{t=0}f_t(m,c_1,e',c_0)}{f(m,c_1,e',c_0)^2}\right\}d\nu(y)\\
&\times f(M=m\mid c_1,e,c_0)f(C_1=c_1\mid e',c_0)f(C_0=c_0)d\nu(m,c_1,c_0)\\
= &\int\limits_{y,m,c_1,c_0} \biggl\{y\frac{\nabla_{t=0}f_t(y,m,c_1,e',c_0)}{f(m,c_1,e',c_0)} - \frac{\nabla_{t=0}f_t(m,c_1,e',c_0)}{f(m,c_1,e',c_0)}\E(Y\mid m,c_1,e',c_0)\biggr.\\
&\times \biggl.f(y\mid m,c_1,e',c_0)\biggr\}f(M=m\mid c_1,e,c_0)f(C_1=c_1\mid e',c_0)f(C_0=c_0)\\
&\times d\nu(y,m,c_1,c_0)\\
= &\int\limits_{y,m,c_1,e^*,c_0} \biggl\{y\frac{\nabla_{t=0}f_t(y,m,c_1,e^*,c_0)}{f(m,c_1,e',c_0)} - \frac{\nabla_{t=0}f_t(m,c_1,e^*,c_0)}{f(m,c_1,e',c_0)}\E(Y\mid m,c_1,e',c_0)\biggr.\\
&\times \biggl. f(y\mid m,c_1,e',c_0)\biggr\}1_{e'}(e^*)f(M=m\mid c_1,e,c_0)f(C_1=c_1\mid e',c_0)f(C_0=c_0)\\
&\times d\nu(y,m,c_1,e^*,c_0)\\
= &\E\left[\frac{1_{e'}(E)f(M\mid C_1,e',C_0)f(C_1\mid e',C_0)f(C_0)}{f(Y,M,C_1,E,C_0)}\left\{Y\frac{\nabla_{t=0}f_t(Y,M,C_1,E,C_0)}{f(M,C_1,e',C_0)}\right.\right.\\
&\left.\left. - f(Y\mid M,C_1,e',C_0)\frac{\nabla_{t=0}f_t(M,C_1,E,C_0)}{f(M,C_1,e',C_0)}B(M,C_1,e',C_0)\right\}\right]\\
= &\E\left[\frac{1_{e'}(E)f(M\mid C_1,e,C_0)f(C_1\mid e',C_0)f(C_0)}{f(Y,M,C_1,E,C_0)f(M,C_1,e',C_0)}\left\{Y\nabla_{t=0}f_t(Y,M,C_1,E,C_0)\right.\right.\\
& - [\nabla_{t=0}f_t(Y,M,C_1,E,C_0) - f(M,C_1,e',C_0)\nabla_{t=0}f_t(Y\mid M,C_1,E,C_0)]\\
&\times \biggl.\left.B(M,C_1,e',C_0)\right\}\biggr]\\
= &\E\biggl[\frac{\nabla_{t=0}f_t(Y,M,C_1,E,C_0)}{f(Y,M,C_1,E,C_0)}\times\frac{1_{e'}(E)f(M\mid C_1,e,C_0)}{f(M\mid C_1,e',C_0)f(E=e'\mid C_0)}\{Y\biggr.\\
&\biggl. - B(M,C_1,e',C_0)\}\biggr] + \int\limits_{m,c_1,c_0}f(c_1\mid e',c_0)f(m\mid c_1,e,c_0)f(c_0)\\
&\times\nabla_{t=0}\left\{\int_y f_t(y\mid m,c_1,e',c_0)d\nu(y)\right\}B(m,c_1,e',c_0)d\nu(m,c_1,c_0)\\
= &\E\left[U_O\frac{1_{e'}(E)f(M\mid C_1,e,C_0)}{f(M\mid C_1,e',C_0)f(E=e'\mid C_0)}\{Y - B(M,C_1,e',C_0)\}\right],\\
\\
(2) = &\int\limits_{m,c_1,c_0}\E(Y\mid m,c_1,e',c_0)\left\{\frac{\nabla_{t=0}f_t(m,c_1,e,c_0)}{f(c_1,e,c_0)} - \frac{\nabla_{t=0}f_t(c_1,e,c_0)f(m,c_1,e,c_0)}{f(c_1,e,c_0)^2}\right\}\\
&\times f(c_1\mid e',c_0)f(c_0)d\nu(m,c_1,c_0)\\
= &\int\limits_{m,c_1,c_0}\E(Y\mid m,c_1,e',c_0)\frac{\nabla_{t=0}f_t(m,c_1,e,c_0)}{f(c_1,e,c_0)}f(c_1\mid e',c_0)f(c_0)d\nu(m,c_1,c_0)\\
&- \int\limits_{c_1,c_0}\frac{\nabla_{t=0}f_t(c_1,e,c_0)}{f(c_1,e,c_0)}\E(\E(Y\mid m,c_1,e',c_0)\mid c_1,e,c_0)f(c_1\mid e',c_0)f(c_0)d\nu(c_1,c_0)\\
= &\int\limits_{m,c_1,c_0}\frac{f(c_1\mid e',c_0)f(c_0)}{f(c_1,e,c_0)}\bigl\{\nabla_{t=0}f_t(m,c_1,e,c_0)\E(Y\mid m,c_1,e',c_0)\\
&-\nabla_{t=0}f_t(c_1,e,c_0)f(m\mid c_1,e,c_0)B'(c_1,e',e,c_0)\bigr\}d\nu(m,c_1,c_0)\\
= &\int\limits_{m,c_1,c_0}\frac{f(c_1\mid e',c_0)}{f(c_1\mid e,c_0)f(e\mid c_0)}\bigl\{\nabla_{t=0}f_t(m,c_1,e,c_0)\E(Y\mid m,c_1,e',c_0)\bigr.\\
&- \bigl.\left[\nabla_{t=0}f_t(m,c_1,e,c_0) - f(c_1,e,c_0)\nabla_{t=0}f_t(m\mid c_1,e,c_0)\right]B'(c_1,e',e,c_0)\bigr\}\\
&\times d\nu(m,c_1,c_0)\\
= &\int\limits_{m,c_1,c_0}\nabla_{t=0}f_t(m,c_1,e,c_0)\frac{f(c_1\mid e',c_0)}{f(c_1\mid e,c_0)f(e\mid c_0)}\left\{E(Y\mid m,c_1,e',c_0) - B'(c_1,e',e,c_0)\right\}\\
&\times d\nu(m,c_1,c_0) + \int\limits_{c_1,c_0}f(c_1\mid e',c_0)f(c_0)\nabla_{t=0}\int_m f_t(m\mid c_1,e,c_0)d\nu(m)B'(c_1,e',e,c_0)\\
&\times d\nu(c_1,c_0)\\
= &\int\limits_{m,c_1,c_0}\biggl\{\int_y f(y\mid m,c_1,e,c_0)d\nu(y)\nabla_{t=0}f_t(m,c_1,e,c_0)\biggr.\\
&\biggl.+ \nabla_{t=0}\int_y f_t(y\mid m,c_1,e,c_0)d\nu(y)f(m,c_1,e,c_0)\biggr\}\frac{f(c_1\mid e',c_0)}{f(c_1\mid e,c_0)f(e\mid c_0)}\\
&\times \left\{\E(Y\mid m,c_1,e',c_0) - B'(c_1,e',e,c_0)\right\}d\nu(m,c_1,c_0)\\
= &\int\limits_{y,m,c_1,c_0}\nabla_{t=0}f_t(y,m,c_1,e,c_0)\frac{f(c_1\mid e',c_0)}{f(c_1\mid e,c_0)f(e\mid c_0)}\\
&\times\left\{\E(Y\mid m,c_1,e',c_0) - B'(c_1,e',e,c_0)\right\}d\nu(y,m,c_1,c_0)\\
= &\int\limits_{y,m,c_1,e^*,c_0}\nabla_{t=0}f_t(y,m,c_1,e^*,c_0)\frac{1_e(e^*)f(c_1\mid e',c_0)}{f(c_1\mid e,c_0)f(e\mid c_0)}\\
&\times\left\{\E(Y\mid m,c_1,e',c_0) - B'(c_1,e',e,c_0)\right\}d\nu(y,m,c_1,e^*,c_0)\\
= &\E\left[U_O\frac{1_e(E)f(C_1\mid e',C_0)}{f(C_1\mid e,C_0)f(e\mid C_0)}\left\{\E(Y\mid M,C_1,e',C_0) - B'(C_1,e',e,C_0)\right\}\right],\\
\\
(3) = &\int\limits_{m,c_1,c_0}\E(Y\mid m,c_1,e',c_0)f(m\mid c_1,e,c_0)\left\{\frac{\nabla_{t=0}f_t(c_1,e',c_0)}{f(e',c_0)}\right.\\
&\left. - \frac{\nabla_{t=0}f_t(e',c_0)f(c_1,e',c_0)}{f(e',c_0)^2}\right\} f(c_0)d\nu(m,c_1,c_0)\\
= &\int\limits_{c_1,c_0}\E(\E(Y\mid M,C_1,e',C_0)\mid c_1,e,c_0)\left\{\frac{\nabla_{t=0}f_t(c_1,e',c_0)}{f(e',c_0)}\right.\\
&\left. - \frac{\nabla_{t=0}f_t(e',c_0)}{f(e',c_0)}f(c_1\mid e',c_0)\right\}f(c_0)d\nu(c_1,c_0)\\
= &\int\limits_{c_1,c_0}B'(c_1,e',e,c_0)\frac{\nabla_{t=0}f_t(c_1,e',c_0)}{f(e',c_0)}f(c_0)d\nu(c_1,c_0)\\
&- \int_{c_0}\E(\E(\E(Y\mid M,C_1,e',C_0)\mid C_1,e,C_0)\mid e',c_0)\frac{\nabla_{t=0}f_t(e',c_0)}{f(e',c_0)}f(c_0)d\nu(c_0)\\
= &\int\limits_{c_1,c_0} \frac{f(c_0)}{f(e',c_0)}\left\{\nabla_{t=0}f_t(c_1,e',c_0)B'(c_1,e',e,c_0)\right.\\
&\left. - \nabla_{t=0}f_t(e',c_0)f(c_1\mid e',c_0)B''(e',e,c_0)\right\}d\nu(c_1,c_0)\\
= &\int\limits_{c_1,c_0}\frac{1}{f(e'\mid c_0)}\left\{\nabla_{t=0}f_t(c_1,e',c_0)B'(c_1,e',e,c_0)\right.\\
&\left.- \left[\nabla_{t=0}f_t(c_1,e',c_0) - \nabla_{t=0}f_t(c_1\mid e',c_0)f(e',c_0)\right]B''(e',e,c_0)\right\}d\nu(c_1,c_0)\\
= &\int\limits_{c_1,c_0}\frac{1}{f(e'\mid c_0)}\nabla_{t=0}f_t(c_1,e',c_0)\left\{B'(c_1,e',e,c_0) - B''(e',e,c_0)\right\}d\nu(c_1,c_0)\\
&+ \int_{c_0}f(c_0)\nabla_{t=0}\int_{c_1}f_t(c_1\mid e',c_0)d\nu(c_1)B''(e',e,c_0)d\nu(c_0)\\
= &\int\limits_{c_1,c_0}\frac{1}{f(e'\mid c_0)}\left\{\int\limits_{y,m}f(y,m\mid c_1,e',c_0)d\nu(y,m)\nabla_{t=0}f_t(c_1,e',c_0)\right.\\
&\left.+ \nabla_{t=0}\int\limits_{y,m}f_t(y,m\mid c_1,e',c_0)d\nu(y,m)f(c_1,e',c_0)\right\}\left\{B'(c_1,e',e,c_0) - B''(e',e,c_0)\right\}\\
&\times d\nu(c_1,c_0)\\
= &\int\limits_{y,m,c_1,c_0}\frac{\nabla_{t=0}f_t(y,m,c_1,e',c_0)}{f(e'\mid c_0)}\left\{B'(c_1,e',e,c_0) - B''(e',e,c_0)\right\}d\nu(y,m,c_1,c_0)\\
= &\int\limits_{y,m,c_1,e^*,c_0}\nabla_{t=0}f_t(y,m,c_1,e^*,c_0)\frac{1_{e'}(e^*)}{f(e'\mid c_0)}\left\{B'(c_1,e',e,c_0) - B''(e',e,c_0)\right\}\\
&\times d\nu(y,m,c_1,e^*,c_0)\\
= &\E\left[U_O\frac{1_e(E')}{f(e'\mid C_0)}\left\{B'(c_1,e',e,C_0) - B''(e',e,C_0)\right\}\right],\\
\end{align*}
and
\begin{align*}
(4) = &\int_{c_0}\E(\E(\E(Y\mid M,C_1,e',C_0)\mid C_1,e,C_0)\mid e',C_0)\nabla_{t=0}f_t(c_0)d\nu(c_0) - \beta_0\E U_O\\
= &\int_{c_0}\left\{\int\limits_{y,m,c_1,e^*}f(y,m,c_1,e^*\mid c_0)d\nu(y,m,c_1,e^*)\nabla_{t=0}f_t(c_0)\right.\\
&\left.+ \nabla_{t=0}\int\limits_{y,m,c_1,e^*}f_t(y,m,c_1,e^*\mid c_0)d\nu(y,m,c_1,e^*)f(c_0)\right\}B''(e',e,c_0)d\nu(c_0)\\
& - \E[U_O\beta_0]\\
= &\int\limits_{y,m,c_1,e^*,c_0}\nabla_{t=0}f_t(y,m,c_1,e^*,c_0)B''(e',e,c_0)d\nu(y,m,c_1,e^*,c_0) - \E[U_O\beta_0]\\
= &\E\left[U_O\left\{B''(e',e,C_0) - \beta_0\right\}\right].
\end{align*}
Thus, $\left.\frac{\partial\beta_t}{\partial t}\right\rvert_{t=0} = \E[U_OEIF(O;\beta_0)]$ where
\begin{align*}
EIF(O;\beta_0) = &\frac{1_{e'}(E)f(M\mid e,C_1,C_0)}{f(M\mid e',C_1,C_0)f(e'\mid C_0)}\left\{Y - B(M,C_1,e',C_0)\right\}\\
&+ \frac{1_{e}(E)f(C_1\mid e',C_0)}{f(C_1\mid e,C_0)f(e\mid C_0)}\left\{B(M,C_1,e',C_0) - B'(C_1,e',e,C_0)\right\}\\
&+ \frac{1_{e'}(E)}{f(e'\mid C_0)}\left\{B'(C_1,e',e,C_0) - B''(e',e,C_0)\right\} + \left\{B''(e',e,C_0) - \beta_0\right\},
\end{align*}
so for any regular, asymptotically-linear estimator in $\mathcal{M}_{np}$, $EIF(\beta_0)$ is the corresponding influence function. It is efficient because the model $\mathcal{M}_{np}$ is nonparametric.
\end{proof}

\begin{proof}[Proof of Theorem 2]
Let $\tilde{B}$, $\tilde{\theta}_M = \{\tilde{M}^{ratio}, \tilde{\E}[\tilde{B}(M,C_1,e',C_0)\mid C_1,e,C_0]\}$, $\tilde{\theta}_{C_1} \allowbreak = \allowbreak \{\tilde{C}_1^{ratio}, \allowbreak \tilde{\E}[\allowbreak \tilde{B'}(\allowbreak C_1,\allowbreak e,\allowbreak C_0)\allowbreak \mid \allowbreak e',\allowbreak C_0]\}$, and $\tilde{f}_{E\mid C_0}$ denote limits of estimators that have limits in probability within the working models $B^W$, $\theta_M^W$, $\theta_{C_1}^W$, and $f_{E\mid C_0}^W$. 
\begin{align*}
\E \{EIF(O;\beta_0,& \tilde{B}, \tilde{\theta}_M, \tilde{\theta}_{C_1}, \tilde{f}_{E\mid C_0})\}=\\
\E\left[\int\limits_{m,c_1}\right.&\frac{\tilde{M}^{ratio}}{\tilde{f}(e'\mid C_0)}\left\{B(m,c_1,e',C_0)-\tilde{B}(m,c_1,e',C_0)\right\}f(m\mid c_1,e',C_0)\\
&\times f(c_1\mid e',C_0)f(e'\mid C_0)d\nu(m,c_1)
+ \int\limits_{c_1}\frac{1}{\tilde{C}_1^{ratio}\tilde{f}(e\mid C_0)}\\
&\times\left\{\E\left[\tilde{B}(M,c_1,e',C_0)\mid c_1,e,C_0\right]-\tilde{\E}\left[\tilde{B}(M,c_1,e',C_0)\mid c_1,e,C_0\right]\right\}\\
&\times f(c_1\mid e,C_0)f(e\mid C_0)d\nu(c_1)\\
&+ \frac{f(e'\mid C_0)}{\tilde{f}(e'\mid C_0)}\left\{\E\left[\tilde{\E}\left[\tilde{B}(M,C_1,e',C_0)\mid C_1,e,C_0\right]\mid e',C_0\right]\right.\\
&\left.-\tilde{\E}\left[\tilde{\E}\left[\tilde{B}(M,C_1,e',C_0)\mid C_1,e,C_0\right]\mid e',C_0\right]\right\}\\
&+ \tilde{\E}\left[\tilde{\E}\left[\tilde{B}(M,C_1,e',C_0)\mid C_1,e,C_0\right]\mid e',C_0\right]\\
&-\E\left[\E\left[B(M,C_1,e',C_0)\mid C_1,e,C_0\right]\mid e',C_0\right]\left.\rule{0cm}{.9cm}\right]
\end{align*}
Substituting under (a):
\begin{align*}
\E \{EIF(O;\beta_0,& \tilde{B}, \tilde{\theta}_M, \tilde{\theta}_{C_1}, \tilde{f}_{E\mid C_0})\}=\\
\E\left[\int\limits_{m,c_1}\right.&\left\{B(m,c_1,e',C_0)-\tilde{B}(m,c_1,e',C_0)\right\}f(m\mid c_1,e,C_0)f(c_1\mid e',C_0)d\nu(m,c_1)\\
+ &\left\{\E\left[\E\left[\tilde{B}(M,C_1,e',C_0)\mid C_1,e,C_0\right]\mid e',C_0\right]\right.\\
&-\left.\tilde{\E}\left[\E\left[\tilde{B}(M,C_1,e',C_0)\mid C_1,e,C_0\right]\mid e',C_0\right]\right\}\\
+ &\tilde{\E}\left[\E\left[\tilde{B}(M,C_1,e',C_0)\mid C_1,e,C_0\right]\mid e',C_0\right]\\
&-\E\left[\E\left[B(M,C_1,e',C_0)\mid C_1,e,C_0\right]\mid e',C_0\right]\left.\rule{0cm}{.9cm}\right]\\
=0&
\end{align*}
Substituting under (b):
\begin{align*}
\E \{EIF(O;\beta_0,& \tilde{B}, \tilde{\theta}_M, \tilde{\theta}_{C_1}, \tilde{f}_{E\mid C_0})\}=\\
\int\limits_{c_1}&\left\{\E\left[B(M,c_1,e',C_0)\mid c_1,e,C_0\right]-\tilde{\E}\left[B(M,c_1,e',C_0)\mid c_1,e,C_0\right]\right\}\\
&\times f(c_1\mid e',C_0)d\nu(c_1)
+\E\left[\tilde{\E}\left[B(M,C_1,e',C_0)\mid C_1,e,C_0\right]\mid e',C_0\right]\\
&-\E\left[\E\left[B(M,C_1,e',C_0)\mid C_1,e,C_0\right]\mid e',C_0\right]\left.\rule{0cm}{.9cm}\right]\\
=0&
\end{align*}
Substituting under (c):\\
$\E \left\{EIF(O;\beta_0, \tilde{B}, \tilde{\theta}_M, \tilde{\theta}_{C_1}, \tilde{f}_{E\mid C_0})\right\}=0$, trivially.

Thus, $\hat{\beta}_{mr}$ can be shown to be asymptotically normal centered at $\beta_0$ under each of these scenarios using a Taylor expansion of $\mathbb{P}_n EIF(\hat{\beta}_{mr}, \hat{B}, \hat{\theta}_M, \hat{\theta}_{C_1}, \hat{f}_{E\mid C_0})$ and applying the central limit theorem to $n^{-1/2}\sum_i EIF(O_i;\beta_0, B^*, \theta_M^*, \theta_{C_1}^*, f_{E\mid C_0}^*)$.
\end{proof}

\newpage

\section{Plot comparing estimators in PEPFAR analysis}

\begin{figure}[h]
\centering
\includegraphics[scale=.75]{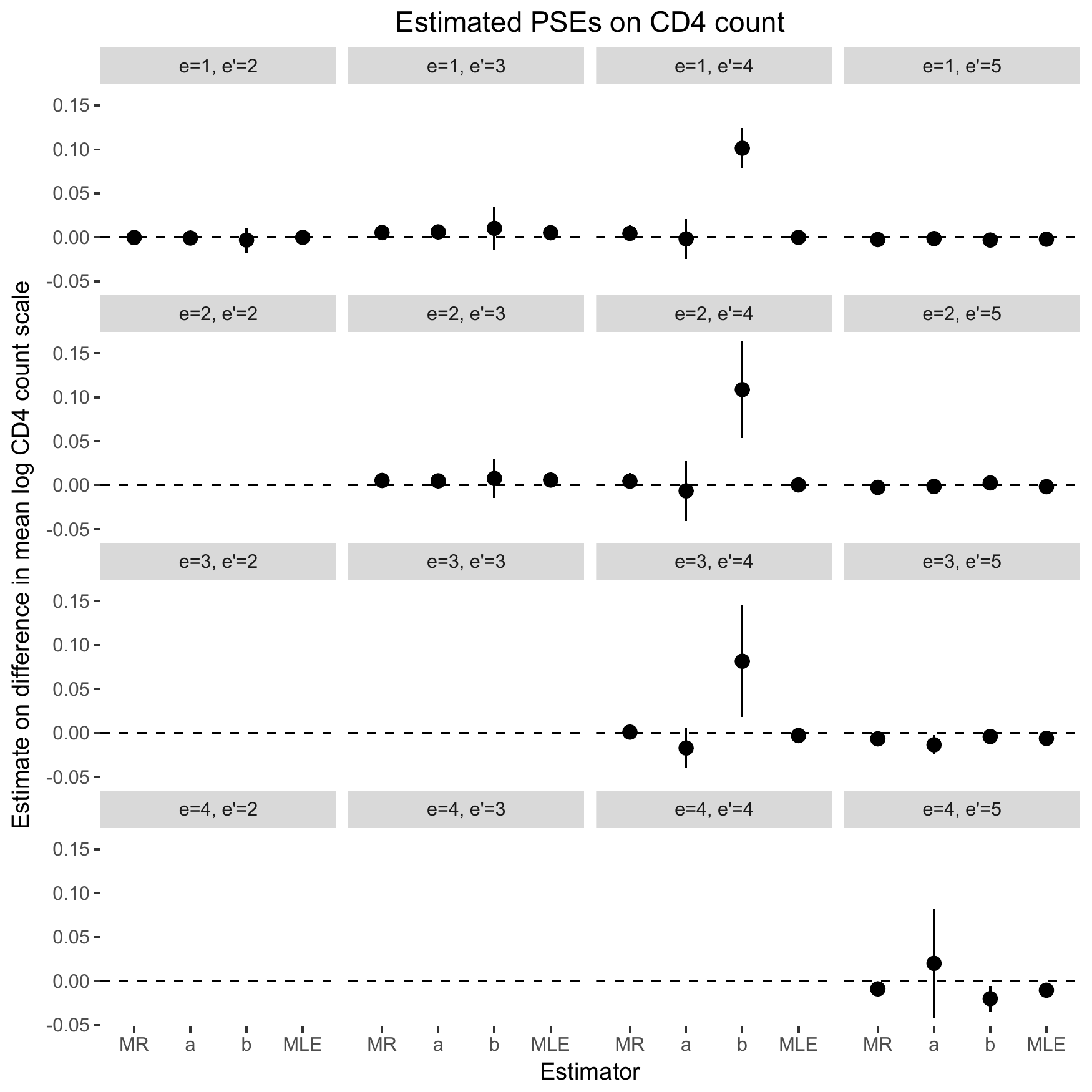}
\caption{$\mathcal{P}_{EMY}$ path specific effects on CD4 count. The plot in each cell represents estimates for the effect with comparison-level treatment, $e$, equal to the first index of the cell and baseline-level treatment, $e'$ equal to the second index of the cell. That is, comparison level treatment varies across rows and baseline level treatment varies across columns. Within each plot, the dots and vertical bars represent point estimates using the four estimators and their corresponding bootstrap confidence intervals.\label{fig:comparison}}
\end{figure}

\bibliographystyle{apalike}

\bibliography{references}

\end{document}